\PassOptionsToPackage{unicode}{hyperref}
\PassOptionsToPackage{hyphens}{url}
\documentclass[
  11pt]{article}
\usepackage{xcolor}
\usepackage[margin=1in]{geometry}
\usepackage{amsmath,amssymb}
\usepackage{iftex}
\ifPDFTeX
  \usepackage[T1]{fontenc}
  \usepackage[utf8]{inputenc}
  \usepackage{textcomp} % provide euro and other symbols
\else % if luatex or xetex
  \usepackage{unicode-math} % this also loads fontspec
  \defaultfontfeatures{Scale=MatchLowercase}
  \defaultfontfeatures[\rmfamily]{Ligatures=TeX,Scale=1}
\fi
\usepackage{lmodern}
\ifPDFTeX\else
\fi
\IfFileExists{upquote.sty}{\usepackage{upquote}}{}
\IfFileExists{microtype.sty}{% use microtype if available
  \usepackage[]{microtype}
  \UseMicrotypeSet[protrusion]{basicmath} % disable protrusion for tt fonts
}{}
\makeatletter
\@ifundefined{KOMAClassName}{% if non-KOMA class
  \IfFileExists{parskip.sty}{%
    \usepackage{parskip}
  }{% else
    \setlength{\parindent}{0pt}
    \setlength{\parskip}{6pt plus 2pt minus 1pt}}
}{% if KOMA class
  \KOMAoptions{parskip=half}}
\makeatother
\usepackage{longtable,booktabs,array}
\usepackage{calc} % for calculating minipage widths
\usepackage{etoolbox}
\makeatletter
\patchcmd\longtable{\par}{\if@noskipsec\mbox{}\fi\par}{}{}
\makeatother
\IfFileExists{footnotehyper.sty}{\usepackage{footnotehyper}}{\usepackage{footnote}}
\makesavenoteenv{longtable}
\usepackage{newunicodechar}
\usepackage{textcomp}
\newunicodechar{°}{\ensuremath{^\circ}}
\newunicodechar{−}{\ensuremath{-}}
\newunicodechar{χ}{\ensuremath{\chi}}
\newunicodechar{ψ}{\ensuremath{\psi}}
\newunicodechar{₀}{\ensuremath{_0}}
\newunicodechar{×}{\ensuremath{\times}}
\newunicodechar{š}{\v{s}}
\usepackage{bookmark}
\IfFileExists{xurl.sty}{\usepackage{xurl}}{} % add URL line breaks if available
\makeatletter
\@ifundefined{xmpquote}{}{}
\makeatother
\hypersetup{
  pdftitle={When does the head leave? Shared timing and a movement clock in 360° video viewing},
  hidelinks,
  pdfcreator={LaTeX via pandoc}}

\title{When does the head leave? Shared timing and a movement clock in
360° video viewing}
\author{Arman Nik Khah\(^1\), Arvin Bahreini\(^2\), Ravi Prakash\(^1\)\\
\(^1\)University of Texas at Dallas, \(^2\)University of Rochester}
\date{}

\begin{document}
\maketitle

\subsection{Abstract}

A 360° streaming system only needs high quality where the viewer faces,
so knowing \emph{when} a head will turn away matters as much as where it
will go. Across four public head-tracking datasets, two used to fix and
validate the method and two held out for testing, three patterns emerge
on the held-out sets. First, viewers facing the same region turn away
within a second of one another far more often than time-shifted traces
allow. That estimate is about five times the pairwise threshold for a
video timing one departure in ten, mapping to roughly one and a third to
over twice that one-in-ten share. The excess co-departing fraction is
0.0422, with a 90\% interval from 0.0279 to 0.0565. Second, time since
the last large head movement improves held-out departure prediction
beyond video and viewer effects. It adds 0.0793 nats per departure, a
predictive dependence rather than proof of an internal timer. Third, on
the eye-tracked test set, gaze crosses the region boundary \emph{after}
the head starts moving, lagging head onset by 0.30 s. Because headsets
record continuous angles, a departure exists only after a rule marks
when a settled hold ends once the head swings past a speed and angle
threshold. Before touching the test sets, we picked one primary setting
for the headline estimates above and tested 18 settings on synthetic
traces with planted answers, requiring at least nine passing settings
for robustness. Fifteen settings passed for the clock and all 15 valid
settings confirm its gain; 8 passed for synchrony and all 8 valid
settings agree, one short of the nine-setting bar; and 6 of 7
dual-validated settings agree on gaze lag. Synchrony validation drops
from 17 or 18 passing settings under guessed noise to 8 under measured
noise because zero-tolerance rules reject a small baseline offset,
consistently positive at one dataset size, as datasets grow.

\subsection{1. Introduction}

Put on a headset, start a 360° video, and watch what your head does. It
settles on a face or a moving car, holds there for a few seconds, then
swings away to something else. A streaming system cares about that swing
more than anything else the viewer does. Only the part of the sphere the
viewer is facing needs to arrive in high quality, so a system that knew
when the head would leave could fetch the next region in time. What
decides \emph{when}?

Three suspects stand out. The video could time the departure when a door
opens, a sound starts, and every viewer looks over at once. The viewer
could time it, because some people are restless and some are patient
whatever plays. Or the head could carry its own rhythm, a clock that
starts at the last large movement, much as the eyes wait a fraction of a
second between rapid eye jumps, or \emph{saccades}. Each suspect makes a
testable prediction. If the video times departures, viewers looking at
the same place in the sphere should leave it together. If a movement
clock times them, the chance of leaving should change with the time
since the last large movement, even after accounting for the video and
the viewer.

Both predictions are easy to state and hard to test. First, a headset
records continuous head angles rather than discrete events; a
\emph{departure} exists only after a segmentation rule, the
\emph{detector}, marks where one settled hold near a direction, a
\emph{dwell}, ends and the next begins once the head exceeds a speed
threshold after holding within an angular radius for a minimum duration.
Second, testing whether viewers leave a region together requires
identifying who occupies that region right before a departure.
Conditioning that reference group on pre-departure presence drops
viewers who just walked out, which biases timestamp-shifted null
controls downward and hides real shared timing. Shifting each viewer's
entire trace intact removes that bias.

Tested on two held-out datasets after locking the method on two
calibration datasets, three behavioral findings emerge. First, viewers
facing the same region leave it together within a second of one another
at about five times the pairwise threshold for a video timing one
departure in ten. Second, the time since the head's last large movement
improves held-out prediction of the next departure beyond video and
viewer effects. Third, on the held-out dataset with eye tracking, a
viewer's gaze crosses the region boundary \emph{after} the head begins
to turn, not before.

To ensure that these patterns do not depend on a single threshold
choice, we selected a primary setting on the calibration sets, requiring
a 60°/s peak speed after a 0.5 s hold within a 30° radius, and filtered
an eighteen-setting grid against synthetic head traces with planted
answers, requiring at least nine passing settings to call any finding
robust across definitions. Fifteen settings passed the clock check and
all fifteen confirm the movement clock; eight passed the synchrony check
and all eight confirm shared departure timing, one short of the
nine-setting robustness bar; and six of seven dual-validated settings
confirm that gaze crosses after head onset.

We contribute (i) held-out measurements of shared departure timing, a
predictive head-movement clock, and eye-head boundary crossing on 360°
head-tracking data, using a discrete-time hazard model that separates
the clock from the video and the viewer; (ii) a demonstration that
conditioning a coincidence test on a state altered by the counted events
makes the obvious time-shift null negatively biased, along with the
whole-trace shift that fixes it; (iii) a demonstration that
zero-tolerance synthetic detector validation breaks as sample size grows
under realistic noise, and how an equivalence margin repairs it; and
(iv) a practical log of timestamp and coordinate traps in these public
datasets.

\subsection{2. Related work}

\textbf{Shared attention in video.} Viewers' gaze moving together is
well documented. In flat video, where people look and how often they
shift their gaze line up across viewers beyond time-shift and
phase-randomised controls (Madsen and Parra 2022; Shepherd et al.~2010;
Wang et al.~2012). In 360° video, viewing direction is correlated across
viewers (Farmer et al.~2021). A head departure is not a gaze shift, and
we did not find a test of whether viewers who share a place \emph{leave}
it together beyond a whole-trace time-shift null; our proof that
timestamp-only shifts are negatively biased when the reference group is
conditioned on pre-departure presence connects viewer synchrony to
state-dependent conditioning traps in epidemiology and economics
(Schmertmann, Assunção and Potter 2010; Miller and Sanjurjo 2018). We
claim the test and its result, not the phenomenon.

\textbf{Timing of eye and head movements.} The time since the last
saccade has been modelled with a survival model in virtual reality
(Rolff et al.~2023); we ask the head's version of that question. In gaze
shifts toward targets, the head routinely starts after the eyes, while
shifts beyond 15° recruit the head and torso to carry the gaze across
wider angles (Sidenmark and Gellersen 2020).

\textbf{Detectors and validation.} Event detectors for eye movements are
known to change results with their thresholds (Andersson et al.~2017).
Checking that an analysis recovers a planted answer before trusting it
on real data is named practice in computational modelling and simulation
studies (Wilson and Collins 2019; Gelman et al.~2020; Siepe et
al.~2024).

\subsection{3. Data}

\subsubsection{3.1 Four datasets, calibration and test}

We use four public datasets of head tracking recorded while people
watched 360° video in a headset (Table 1). Two calibration sets fix
every value the method needs from real heads, such as how noisy trackers
are, and nothing reported as a finding comes from them. Two test sets
were downloaded only after the method was fixed, and every headline
result comes from them.

{\def\LTcaptype{none} % do not increment counter
\begin{longtable}[]{@{}
  >{\raggedright\arraybackslash}p{(\linewidth - 10\tabcolsep) * \real{0.1667}}
  >{\raggedright\arraybackslash}p{(\linewidth - 10\tabcolsep) * \real{0.1667}}
  >{\raggedright\arraybackslash}p{(\linewidth - 10\tabcolsep) * \real{0.1667}}
  >{\raggedright\arraybackslash}p{(\linewidth - 10\tabcolsep) * \real{0.1667}}
  >{\raggedright\arraybackslash}p{(\linewidth - 10\tabcolsep) * \real{0.1667}}
  >{\raggedright\arraybackslash}p{(\linewidth - 10\tabcolsep) * \real{0.1667}}@{}}
\toprule\noalign{}
\begin{minipage}[b]{\linewidth}\raggedright
Role
\end{minipage} & \begin{minipage}[b]{\linewidth}\raggedright
Dataset
\end{minipage} & \begin{minipage}[b]{\linewidth}\raggedright
Viewers per video
\end{minipage} & \begin{minipage}[b]{\linewidth}\raggedright
Videos
\end{minipage} & \begin{minipage}[b]{\linewidth}\raggedright
Usable length
\end{minipage} & \begin{minipage}[b]{\linewidth}\raggedright
Tracking
\end{minipage} \\
\midrule\noalign{}
\endhead
\bottomrule\noalign{}
\endlastfoot
Calibration & Nasrabadi et al., MMSys 2019 & 30 & 28 & 60 s & head, 60
Hz \\
Calibration & Jin et al., ACM MM 2022 & 92 to 100 (measured) & 27 & 60 s
& head and gaze, 120 Hz \\
Test & Wu et al., MMSys 2017, Experiment 1 & 48 & 9 & 164 to 655 s &
head, about 100 Hz \\
Test & Panonut360 (Xu et al., MMSys 2024) & 50 & 15 & 140 to 412 s &
head and eye, 120 Hz \\
\end{longtable}
}

\emph{Table 1.} The datasets. Wu's Experiment 2 had a memory task and is
not used.

Each dataset fills a specific role in the design. Nasrabadi's
camera-motion labels give us positive controls, Jin's gaze lets us check
what a head departure is, and Wu and Panonut360 have the long clips a
clock analysis needs. We read tracking files only, with no
questionnaire, per-participant metadata or video, and no linkage to
other data. The unit of analysis is the video; intervals are computed
across videos unless stated otherwise.

\subsubsection{3.2 Preprocessing}

Every dataset goes through the same pipeline. Each head sample becomes a
unit vector in one coordinate frame using the mapping in its paper, and
is resampled to 30 Hz by spherical interpolation along the shortest arc.
A tracking gap longer than 0.2 s splits a trace and censors the dwell
that spans it, ending that dwell at the gap without counting a
departure. Angular speed is the norm of a 7-sample quadratic smoothing
derivative, Savitzky-Golay order 2, with its radial part removed. We
exclude the first 15 s of each video and drop any video with fewer than
10 usable viewers. All four datasets pass our kinematic sanity checks,
which reject any dataset whose 99.9th percentile of head speed reaches
600°/s or whose median horizontal viewing angle, or \emph{yaw}, over
each video's first 2 s lies more than 30° from the front. For Wu the yaw
check is weak, because Wu passes on the forward axis, a front its paper
does not document, and would fail against the paper's literal reading.

\subsubsection{3.3 Reading the test sets}

Both test sets disagreed with their papers, and anyone reusing them will
meet the same problems. Wu et al.'s archive numbers its nine Experiment
1 videos 0 to 8, where the paper's table numbers them 1 to 9; a loader
written from the paper drops one video in nine. Panonut360 names its
columns differently from its paper; the archive's own column guide
settles the mapping.

Wu's video timestamp, PlaybackTime, steps backwards somewhere in 415 of
its 432 traces. Replacing it with the wall-clock timestamp looks like
the obvious fix, and fails. Tied to each trace's first PlaybackTime, the
wall-clock timestamp drifts ahead of the video because many files start
with a stale PlaybackTime row. By the end of a trace, that drift reaches
a median of 1.08 s and a maximum of 34 s, wider than our one-second
synchrony window. Instead, we used PlaybackTime, dropped the stale
startup rows, and removed any row where PlaybackTime stepped backwards.
After startup, PlaybackTime tracks the wall-clock timestamp within 0.03
s, except in a tenth of the traces where video playback stalls while the
wall-clock timestamp keeps running, ending 7.5 s or more apart. Shared
timing on Wu clears the threshold under the wall-clock timestamp as
well. That alternative reading is 0.0377, with a 90\% interval from
0.0252 to 0.0503.

Viewers depart at 3.98 per viewer-minute on Wu and 5.01 on Panonut360 at
the primary setting. In total, Wu provides 432 traces with a
99.9th-percentile head speed of 227.6°/s, and Panonut360 provides 749
traces across 15 videos with 49 or 50 viewers each and a
99.9th-percentile speed of 258.6°/s.

\subsection{4. Method}

\subsubsection{4.1 Detecting dwells and departures}

Our segmentation rule turns continuous head traces into \emph{dwells},
stretches where the head stays near one direction, and
\emph{departures}, fast rotations that carry it away. Any run of samples
faster than 20°/s counts as a \emph{movement}. A dwell's clock starts
when the movement that brought the head there ends. Once the head
remains within an angular radius \texttt{A} of that arrival direction
for more than a minimum hold duration \texttt{d\_min}, the dwell becomes
\emph{established}; its \emph{anchor} is the mean settled direction, and
only an established dwell can end in a departure. A movement from an
established dwell is a \emph{departure} if it reaches a peak speed
\texttt{v\_peak} and lands more than \texttt{A} from the anchor. A
movement that lands within \texttt{A} is an \emph{adjustment} and
updates the anchor. If a movement reaches \texttt{v\_peak} and travels
at least 20° but lands back inside \texttt{A}, it censors the dwell
without counting a departure and restarts the clock. Thus, the clock
measures \textbf{time since the head last made a large movement}.

We report every headline measurement at one \emph{primary} setting
requiring a 60°/s peak speed, a 30° radius, and a 0.5 s minimum hold. To
check sensitivity across threshold definitions, we cross
\texttt{v\_peak} values of 40, 60, or 90°/s, radii \texttt{A} of 20, 30,
or 45°, and minimum holds \texttt{d\_min} of 0.5 or 1 s to form an
18-setting grid. We selected the primary setting on Jin, where gaze is
recorded, as the validated setting whose head departures best aligned
with gaze shifts larger than \texttt{A}. Even there, head and eye
movements often diverge. Across 11,944 head departures at the primary
setting, 56.7\% had no gaze shift of the same size within half a second.
Every claim in this paper is about the head.

\subsubsection{4.2 Measuring shared timing and the co-set shift trap}

To test whether viewers leave the same place together, take a departure
by one viewer at video time t. Its \emph{co-viewers} are the other
viewers of the same video whose heads were within \texttt{A} of the
departing viewer's anchor half a second earlier, scored whenever there
are at least three co-viewers. For that departure, we compute the
fraction of co-viewers who also depart within one second of t. To find
what that fraction would be by chance, we shift each co-viewer's entire
trace, head direction and departures together, by a random offset of up
to 5 s, re-identify who is within \texttt{A} on the shifted traces, and
recompute the fraction of co-viewers who depart within one second of t.
Our synchrony statistic χ is the observed fraction minus the chance
fraction, averaged across departures within each video and then across
videos, with a 90\% t-interval across videos.

\textbf{Why shifting only departure times fails.} An obvious shortcut is
to keep the co-viewers picked on the unshifted traces and randomly shift
only their departure timestamps. That shortcut is negatively biased
enough to hide a real effect. With nothing planted, shifting timestamps
alone produced negative χ values from −0.014 to −0.019 across all 18
settings, whereas a synthetic world where the video times one departure
in ten has an oracle value of +0.015. The bug is a headcount taken after
people have already walked out, because a co-viewer who left the region
in the half second before the count is excluded from the unshifted
co-viewer set even though their departure falls inside the scoring
window, whereas the timestamp-only shift counts such departures freely.
Shifting each co-viewer's whole trace and re-picking the co-viewers
eliminates the bias at 18 of 18 settings. Whenever a coincidence count
selects its reference group by a state that the counted events
themselves alter, holding that group fixed under the null biases the
test toward finding no synchrony (Schmertmann, Assunção and Potter 2010;
Miller and Sanjurjo 2018).

\textbf{The decision threshold.} We set our smallest effect worth
detecting at a video-timed share of ψ = 0.10, meaning the video times
one departure in ten. Because a pair of viewers co-departs at a scene
event only when both the departing viewer and a co-viewer respond to
that event, χ grows with the square of ψ, and between-video spread in ψ
raises average χ at a fixed mean share. To express the ψ = 0.10 cutoff
in observable pair units, we evaluate χ on synthetic traces with ψ =
0.10 under the calibration-measured noise levels and test-set sizes to
obtain the decision threshold χ*. At the primary setting on the pooled
test sets, χ* is 0.00861, against a no-effect baseline of 0.00059 with
nothing planted. Shared timing counts as \emph{present} when χ reaches
χ* and its 90\% lower bound clears the no-effect reading; because χ*
varies by more than a factor of two across unmeasured shared-event rates
and clip lengths, we also test χ against the threshold at the top of the
plausible event-rate range.

\subsubsection{4.3 Modeling the movement clock}

To test whether the time since the last large movement predicts the next
departure, we slice each dwell into 0.25 s steps, up to 15 s, and fit a
discrete-time hazard model with a complementary log-log link and three
additive terms. The \emph{clock} is a smooth curve in log dwell time.
The \emph{viewer} term is a per-viewer random effect that absorbs both
steady traits and single-session state. The \emph{scene} term is a
random effect per video and per second of video time, shared by every
viewer watching that second. We fit the model via the Laplace
approximation, which matched adaptive quadrature closely and credited
every planted term and no unplanted one across eight synthetic worlds.

We evaluate the clock by holding out videos in five folds and comparing
the full model against the same model with a constant clock. Our metric,
the \emph{clock gain}, is the improvement in held-out log-likelihood per
departure, measured in nats. Because a detector that chops up dwells can
manufacture apparent duration dependence on a flat clock, we do not test
the clock gain against zero. In one calibration test, a candidate
detector turned a flat planted clock into a held-out gain of +0.00054,
with a 90\% interval from +0.00006 to +0.00104. We therefore compare the
gain against a synthetic false-positive floor g₀, defined as the largest
90\% upper bound of the held-out gain on flat-clock synthetic data at
that setting over two seeds. At the primary setting, g₀ is 0.000507 nats
per departure. The clock counts as \emph{real} when the gain's 90\%
lower bound, bootstrapped across videos, exceeds g₀, and as \emph{flat}
when its upper bound sits below 1\% of the constant-clock model's
held-out log-likelihood per departure, a ceiling matching the gain of a
strong planted peaked clock. At the primary setting, a planted
rising-and-leveling clock always cleared g₀ and a planted flat one never
did.

\subsubsection{4.4 Measuring gaze lead at departure}

To test whether the eyes leave a region before the head starts turning,
for each head departure we define \emph{gaze exit} as the first moment
the viewer's gaze moves more than \texttt{A} from the anchor and stays
outside \texttt{A} for 100 ms, searching from 1.0 s before head onset,
clipped at the dwell's start, to 0.5 s after. The \emph{gaze lead} is
head onset time minus gaze exit time, so a positive lead means the eyes
crossed the boundary before the head began to move. We average the
per-video median lead across videos on Panonut360, the only test set
with eye tracking, and require a lead of 0.15 s or more to conclude that
the eyes announce the departure, a judgment call based on the head's
routine lag behind the eye in target shifts (Sidenmark and Gellersen
2020).

\subsubsection{4.5 Filtering the 18-setting grid on synthetic traces}

Before running these estimators on the held-out test sets, we filter the
18-setting detector grid on synthetic head traces with planted answers
and nuisance levels measured on the calibration sets, including
adjustments fast enough to cross the speed thresholds, tracker noise,
smooth pursuit of moving objects, brief dwells, departures whose size
straddles \texttt{A}, and viewer-to-viewer differences in restlessness.
We run each synthetic trace through the detector and compare its
estimate against an \emph{oracle} that computes the same statistic
directly from the planted events.

If a detector setting fails validation for a question, we exclude it for
that question. We call a finding \emph{robust across detector settings}
when the primary setting's verdict holds at 80\% or more of the
validated settings, with at least nine passing; with fewer than nine
passing settings, robustness is not established. At the
calibration-measured noise levels, 15 of 18 settings passed the clock
check and 8 passed the synchrony check, making a robustness claim
possible for the clock and one setting short for synchrony (Section
6.2).

\subsection{5. Results}

\subsubsection{5.1 Viewers who share a place leave it together}

Viewers facing the same region of the sphere turn away within a second
of one another far more often than time-shifted copies of their traces
allow. Shared timing χ reaches 0.0422 on the pooled test sets at the
primary setting, with a 90\% interval from 0.0279 to 0.0565. That
estimate is about five times the one-in-ten threshold χ* of 0.00861 and
seventy times the no-effect baseline of 0.00059. It also clears the
highest threshold across the plausible range of shared event rates,
where χ* reaches 0.0166 at 4 shared events per minute. Both test sets
show the effect individually: Wu gives 0.0410, with a 90\% interval from
0.0287 to 0.0532, and Panonut360 gives 0.0429, with a 90\% interval from
0.0201 to 0.0657. Across the detector grid, all eight validated
synchrony settings agree at 3.6 to 6.2 times their respective
thresholds, one setting short of the nine our robustness rule requires
(Section 6.2). The calibration sets show the same pattern, reaching 2.4
times the pooled threshold on Nasrabadi and 4.6 times on Jin, though
they tuned the method and do not count as confirmatory evidence. On
Nasrabadi's six fixed-camera videos alone, the ratio is 3.6 times
threshold, with a wide interval from −0.0078 to 0.0705 that cannot rule
out camera motion on that small subset.

When viewers leave together, their departures line up within roughly 0.3
to 0.6 s on the generator's map, and are not resolved further. Shifting
co-viewer traces by at most half a second instead of up to five keeps
the shifted traces mostly inside the one-second scoring window and
eliminates nearly all of the excess coincidence, dropping descriptive χ
to 0.0021 on Wu, from 0.0010 to 0.0032, and to 0.0009 on Panonut360,
from −0.0003 to 0.0021. Viewers watching different videos show no excess
coincidence when paired across videos, yielding 0.0060 on Wu, from
−0.0027 to 0.0147, and −0.0002 on Panonut360, from −0.0054 to 0.0050. We
report this cross-video check descriptively because it failed its own
synthetic validation.

In share terms, the pooled point estimate corresponds to the video
timing between 1.3 and over 2 times the 0.10 floor in the synthetic
generator (Section 4.2). At the measured between-video spread in ψ, the
90\% interval's lower end maps below 0.10, with the uncertainty in χ*
applying on top.

\subsubsection{5.2 The time since the last large movement predicts the
next departure}

Knowing how long ago the head last made a large movement improves
held-out prediction of the next departure beyond video and viewer
effects. Adding the movement clock raises held-out log-likelihood by
0.0793 nats per departure on the pooled test sets at the primary
setting, with a 90\% interval from 0.0698 to 0.0892, well above the
0.0005 false-positive floor g₀. Both test sets show the gain
individually, with Wu at 0.0965 nats per departure, from 0.0865 to
0.1116, and Panonut360 at 0.0740, from 0.0620 to 0.0864. All 15
validated clock settings confirm the effect, establishing robustness
across detector settings. Across pooled dwells, the empirical departure
rate falls as a hold lengthens, so a dwell that has lasted 10 s is about
a third as likely to end in the next step as a dwell at 1 s. At the
primary setting, the Weibull slope is −0.481 on Wu, from −0.503 to
−0.458, and −0.439 on Panonut360, from −0.455 to −0.422, compared with
−0.041 to 0.043 on flat-clock synthetic traces. We fitted these Weibull
slopes across all 15 validated settings, where zero marks a flat hazard,
as a descriptive check after the main results were known.

Neither the held-out gain nor the negative Weibull slope proves that an
individual dwell's internal timer slows down. When some viewers or
individual dwells are more restless than others, restless dwells end
early and leave patient ones behind, producing a falling aggregate curve
and predictive duration dependence even when every individual dwell has
a constant hazard (Vaupel et al.~1979).

\subsubsection{5.3 Gaze crosses the boundary after the head starts
moving}

On Panonut360, a viewer's gaze crosses the region boundary about 0.30 s
\emph{after} the head starts turning. The per-video median gaze lead at
the primary setting is −0.296 s across 16,489 departures in 15 videos,
with a 90\% interval from −0.303 to −0.288, ruling out our 0.15 s
eye-lead threshold. Gaze crosses the boundary first in only 16.6\% of
departures, and 26.6\% have no gaze exit within the search window. Jin
showed the same order with a median lead of −0.128 s. Just before head
onset, head speed predicts an upcoming departure modestly while gaze
distance from the anchor, or \emph{gaze eccentricity}, does not, with
areas under the ROC curve of 0.61 and 0.48.

Gaze crosses before head onset only when the region radius shrinks to
20°, where the single validated setting at 40°/s peak speed and a 1 s
minimum hold gives a positive lead of +0.317 s. Six of the seven
settings that passed both the synchrony and clock checks agree that gaze
crosses after head onset, and with seven dual-validated settings,
robustness across definitions is not established.

\subsection{6. Discussion}

\subsubsection{6.1 Two regularities, possibly one}

Viewers who share a place in a 360° video leave it with video-locked
timing, and the time since the head's last large movement predicts the
next departure. Although we measured these two regularities separately,
a single mechanism could produce both, because viewers who arrive
together at a salient moment and then run the same movement clock will
leave together later with no new event in the video. Our three-term
model can in principle apportion departures among clock, viewer, and
scene, but at the test sets' sizes its synthetic recovery test failed,
leaving that arrival echo unseparated. Likewise, because a scene cut, a
camera pan, or a sudden sound pulls heads together just as a story beat
does, shared timing shows that departures lock to the video timeline
without isolating which feature of the video drove them.

Within-session drift is too small to explain the movement clock's
predictive gain. The measured gain of 0.079 nats per departure is forty
times the largest synthetic session-drift gain and exceeds every planted
rising clock, which gained 0.037 to 0.053. On synthetic traces with a
flat clock, adding session drift produces held-out gains of 0.0003 to
0.0020 nats per departure whose lower bounds through the detector stay
below g₀, leaving between-dwell frailty heterogeneity (Vaupel et
al.~1979) as the main alternative to an internal timer.

The gaze result looks like a reversal of the familiar eye-first order of
eye-head coordination, and it probably is not one. The eyes may well
start moving first while a shift wider than their own range crosses a
boundary 20° to 45° away only once head rotation carries the gaze across
it. The single 20° setting flips to a positive lead of +0.317 s, fitting
that geometric reading, though a single setting that also differs in
speed threshold and minimum hold cannot show that the lead tracks region
size; we did not test whether it does. We did not test whether these
regularities improve viewport prediction.

\subsubsection{6.2 Why zero-tolerance synthetic validation shrinks with
sample size}

A zero-tolerance synthetic validation rule breaks as datasets grow
because larger samples shrink confidence intervals around tiny, benign
baseline offsets. When a validation rule requires the no-effect
confidence interval to contain the oracle's exact value at every seed
and dataset size, even a tiny offset causes systematic failures once the
sample is large enough. Our synchrony check passed at 17 or 18 of 18
settings under guessed noise levels, but dropped to 8 of 18 under
calibration-measured noise levels. At Wu's sample size with nothing
planted, the synchrony statistic reads +0.00060 above the oracle on
average and is positive across all 18 of 18 settings, whereas at
Panonut360's size its sign changes between draws, reading −0.00046 on
the first. As intervals tighten across seeds and sizes, that offset
drives the pass count from 18 of 18 settings at one small size down to 8
of 18 across both test sizes. The runs differ in more than size, so that
drop is a suggestive pattern rather than a dose-response curve. Because
the offset is about 0.001 while the measured χ is 0.042, the offset is
forty times too small to explain the held-out effect.

Two defensible coordinate readings of the same calibration dataset also
flip which result meets the robustness bar. Jin stores head direction
both as a 2-D image point, which we registered before either count
existed, and as a 3-D vector in the headset's frame; switching to the
3-D vector shifts the measured noise levels so the synchrony check
passes at 15 settings instead of 8 and the clock check passes at 7
instead of 15. Meanwhile, the clock check can confirm a clock far more
easily than it can confirm the absence of one. A planted
rising-and-leveling clock clears g₀ at every seed and size at the
primary setting, whereas a planted peaked clock clears g₀ at only 4 of
18 settings, so validation could have vouched for a flat clock at only 2
of the 15 settings that passed.

We recommend four repairs for synthetic detector validation: first, give
each check an equivalence margin for the no-effect reading, fixed before
noise levels are measured, so larger samples do not penalize tiny benign
offsets; second, power the validation check for the measured noise
levels, just as one powers the primary test; third, report pass counts
alongside the sign and spread of their baseline offsets, separating
coin-edge misses from large biases; and fourth, settle how each dataset
is read using a physical invariant before its noise levels feed
validation. On Jin, applying each collection's own clock, where the
third video starts at 120 s, and its own image size of 1280 × 720, 1920
× 1080, or 2560 × 1440 places viewers at +4° from the video front at
start, whereas a naive single-resolution read with a running clock
places them 58° off-center.

\subsubsection{6.3 Limitations}

The synthetic worlds are not the real one. Real dwell lengths vary far
more than our generator allows, as a censored gamma fit gives a shape
parameter of 0.044 against the generator's floor of 0.45, and that same
generator calibrates g₀ and χ*. Real heads are also continuously active.
In 98\% of one-second windows inside a dwell the head moves faster than
three times tracker noise, and one departure in seven starts within 0.3
s of a small corrective movement. Finally, confirmation rests on two
test sets of 9 and 15 videos, where Wu alone has limited power for the
smallest effect and χ* remains uncertain by more than a factor of two.

\subsection{7. Data, ethics and reproducibility}

\textbf{Reproducibility.} Apart from the dataset sizes in Table 1, taken
from each dataset's paper, every reported number but three traces to a
committed run that records its command, code version and environment;
the three exceptions, namely the Laplace fit's agreement with
quadrature, the 1\% flat margin, and a strong peak's gain, trace to an
internal calibration note. The main test-set run, rerun from a fresh
checkout of its own commit, gave an identical result apart from
wall-clock fields.

\subsection{References}

Andersson, R., Larsson, L., Holmqvist, K., Stridh, M., \& Nyström, M.
(2017). One algorithm to rule them all? An evaluation and discussion of
ten eye movement event-detection algorithms. \emph{Behavior Research
Methods, 49}(2), 616--637. https://doi.org/10.3758/s13428-016-0738-9

Farmer, H., Bevan, C., Green, D., Rose, M., Cater, K., \& Stanton
Fraser, D. (2021). Did you see what I saw?: Comparing attentional
synchrony during 360° video viewing in head mounted display and tablets.
\emph{Journal of Experimental Psychology: Applied, 27}(2), 324--337.
https://doi.org/10.1037/xap0000332

Gelman, A., Vehtari, A., Simpson, D., Margossian, C. C., Carpenter, B.,
Yao, Y., Kennedy, L., Gabry, J., Bürkner, P.-C., \& Modrák, M. (2020).
\emph{Bayesian workflow}. arXiv:2011.01808.

Jin, Y., Liu, J., Wang, F., \& Cui, S. (2022). Where are you looking? A
large-scale dataset of head and gaze behavior for 360-degree videos and
a pilot study. In \emph{Proceedings of the 30th ACM International
Conference on Multimedia (MM '22)} (pp.~1025--1034). ACM.
https://doi.org/10.1145/3503161.3548200

Madsen, J., \& Parra, L. C. (2022). Cognitive processing of a common
stimulus synchronizes brains, hearts, and eyes. \emph{PNAS Nexus, 1}(1),
pgac020. https://doi.org/10.1093/pnasnexus/pgac020

Miller, J. B., \& Sanjurjo, A. (2018). Surprised by the hot hand
fallacy? A truth in the law of small numbers. \emph{Econometrica,
86}(6), 2019--2047. https://doi.org/10.3982/ECTA14943

Nasrabadi, A. T., Samiei, A., Mahzari, A., McMahan, R. P., Prakash, R.,
Farias, M. C. Q., \& Carvalho, M. M. (2019). A taxonomy and dataset for
360° videos. In \emph{Proceedings of the 10th ACM Multimedia Systems
Conference (MMSys '19)} (pp.~273--278). ACM.
https://doi.org/10.1145/3304109.3325812

Rolff, T., Schmidt, S., Steinicke, F., \& Frintrop, S. (2023). A deep
learning architecture for egocentric time-to-saccade prediction using
Weibull mixture-models and historic priors. In \emph{Proceedings of the
2023 Symposium on Eye Tracking Research and Applications (ETRA '23)}.
ACM. https://doi.org/10.1145/3588015.3588408

Schmertmann, C. P., Assunção, R. M., \& Potter, J. E. (2010). Knox meets
Cox: Adapting epidemiological space-time statistics to demographic
studies. \emph{Demography, 47}(3), 629--650.
https://doi.org/10.1353/dem.0.0113

Shepherd, S. V., Steckenfinger, S. A., Hasson, U., \& Ghazanfar, A. A.
(2010). Human--monkey gaze correlations reveal convergent and divergent
patterns of movie viewing. \emph{Current Biology, 20}(7), 649--656.
https://doi.org/10.1016/j.cub.2010.02.032

Sidenmark, L., \& Gellersen, H. (2020). Eye, head and torso coordination
during gaze shifts in virtual reality. \emph{ACM Transactions on
Computer-Human Interaction, 27}(1), 1--40.
https://doi.org/10.1145/3361218

Siepe, B. S., Bartoš, F., Morris, T. P., Boulesteix, A.-L., Heck, D. W.,
\& Pawel, S. (2024). Simulation studies for methodological research in
psychology: A standardized template for planning, preregistration, and
reporting. \emph{Psychological Methods}. Advance online publication.
https://doi.org/10.1037/met0000695

Vaupel, J. W., Manton, K. G., \& Stallard, E. (1979). The impact of
heterogeneity in individual frailty on the dynamics of mortality.
\emph{Demography, 16}(3), 439--454. https://doi.org/10.2307/2061224

Wang, H. X., Freeman, J., Merriam, E. P., Hasson, U., \& Heeger, D. J.
(2012). Temporal eye movement strategies during naturalistic viewing.
\emph{Journal of Vision, 12}(1), 16. https://doi.org/10.1167/12.1.16

Wilson, R. C., \& Collins, A. G. E. (2019). Ten simple rules for the
computational modeling of behavioral data. \emph{eLife, 8}, e49547.
https://doi.org/10.7554/eLife.49547

Wu, C., Tan, Z., Wang, Z., \& Yang, S. (2017). A dataset for exploring
user behaviors in VR spherical video streaming. In \emph{Proceedings of
the 8th ACM Multimedia Systems Conference (MMSys '17)} (pp.~193--198).
ACM. https://doi.org/10.1145/3083187.3083210

Xu, Y., Du, J., Wang, J., Ning, Y., Zhou, S., \& Cao, Y. (2024).
Panonut360: A head and eye tracking dataset for panoramic video. In
\emph{Proceedings of the 15th ACM Multimedia Systems Conference (MMSys
'24)} (pp.~319--325). ACM. https://doi.org/10.1145/3625468.3652176

\end{document}